# Self-sustaining Software Systems (S4): Towards Improved Interpretability and Adaptation

Christian Cabrera
Department of Computer Science and Technology, University of Cambridge
United Kingdom
chc79@cam.ac.uk

Andrei Paleyes
Department of Computer Science and Technology, University of Cambridge
United Kingdom
ap2169@cam.ac.uk

Neil D. Lawrence
Department of Computer Science and Technology, University of Cambridge
United Kingdom
ndl21@cam.ac.uk

## ABSTRACT
Software systems impact society at different levels as they pervasively solve real-world problems. Modern software systems are often so sophisticated that their complexity exceeds the limits of human comprehension. These systems must respond to changing goals, dynamic data, unexpected failures, and security threats, among other variable factors in real-world environments. Systems' complexity challenges their interpretability and requires autonomous responses to dynamic changes. Two main research areas explore autonomous systems' responses: evolutionary computing and autonomic computing. Evolutionary computing focuses on software improvement based on iterative modifications to the source code. Autonomic computing focuses on optimising systems' performance by changing their structure, behaviour, or environment variables. Approaches from both areas rely on feedback loops that accumulate knowledge from the system interactions to inform autonomous decision-making. However, this knowledge is often limited, constraining the systems' interpretability and adaptability. This paper proposes a new concept for interpretable and adaptable software systems: self-sustaining software systems (S4). S4 builds knowledge loops between all available knowledge sources that define modern software systems to improve their interpretability and adaptability. This paper introduces and discusses the S4 concept.

## 1 INTRODUCTION
Software systems have become pervasive, impacting our society at different levels [28]. They support our daily activities by satisfying requirements like recommending the content we consume, planning our commutes in a city, or enabling scientific discoveries. Technological advances, the emergence of challenging data requirements, and the difference between isolated development settings and dynamic real-world environments have made systems more complex over time [3, 12]. This growing complexity and the dynamic environments challenge systems' interpretability, maintenance and sustainability [12]. Modern software systems are often so sophisticated that their complexity exceeds the limits of human comprehension and control [23]. Deployed systems must respond to changing goals, variable data, unexpected failures, and security threats, among other variables that emerge from demanding requirements and real-world environments [24]. Such responses must be autonomous because software developers, systems managers, and end-users do not control all systems components [22]. Systems stakeholders must understand the reasons behind the systems' behaviour despite such automation.

Two main research areas explore the development of autonomous systems that respond to dynamic conditions. Evolutionary computing develops optimisation and problem-solving techniques inspired by natural evolution since the 1940s [7]. Genetic improvement optimises existing software based on genetic programming. This process makes small mutations to the system source code, creating new system versions. Each version is benchmarked against the previous one to evaluate the system improvement for a given criteria [17]. Autonomic computing has researched the design and implementation of self-managing, self-adaptive, self-configuration, self-healing, and self-optimisation systems in last decades [16]. This field aims to build systems that respond to variable environments without human intervention. It proposes to equip systems with sensors for environment monitoring, decision functions to reason about perceived changes, and actors to modify systems' structure, behaviour, and environment variables. These elements constitute feedback loops that accumulate knowledge from systems' internal and external interactions. This knowledge drives adaptive decisions [39]. Both research areas are fruitful fields with several successful applications [26, 40]. They have contributed to developing architectures, design patterns, methodologies, algorithms, and tools for developing autonomous systems. However, approaches proposed in these fields are often limited. They are based on processes that work well when optimising simple non-functional requirements (e.g., the number of bugs in the code, the response time of the systems, or the shortest path for an autonomous vehicle) [2, 17]. Modern systems have demanding high-level requirements such as accountability, trustworthiness, transparency, and fairness [6, 31]. Engineers deploy systems in dynamic environments with changing goals and complex uncertainties. Current autonomous systems are limited to runtime representations of goals and fail to model real-life complex phenomena [39]. Their optimisation procedures are usually black-box and hard to explain for sufficiently complex systems [4]. This lack of transparency impacts the end user's ability to interpret the software's behaviour. We argue that the limitations of modern autonomous systems arise from the intrinsic nature of evolutionary and autonomic computing. Current autonomous systems make adaptive decisions mainly based on knowledge from their internal and external interactions, which are encoded in black-box models.

This knowledge is insufficient and constrains the systems' self-sustaining capabilities. Modern and future end-user requirements and the current software development trends demand a shift in the design paradigms of autonomous systems.

This paper introduces the concept of self-sustaining software systems (S4) (Section 2), which leads to improved interpretability of systems leveraging the machine capabilities to self-analyse and self-maintain through shifts in software engineering (SE) practices [3] and advances in artificial intelligence (AI) [36]. S4 empowers users such as clinicians, scientists, and manufacturers and enables the development of complex systems that operate together with humans and remain under their control. The idea behind S4 is to replace the feedback loops that current autonomous systems use with knowledge loops that would enrich both systems' interpretability and adaptability. Such loops integrate and exploit the three sources of knowledge that define modern software systems. The first source is the developers who encode the users' requirements and their expert knowledge into design models. The deployed systems are the second source of knowledge. Their states and internal and external interactions provide knowledge about systems' behaviour. The third source is the knowledge produced by the SE community. Such knowledge provides insights into best practices for systems management based on empirical evidence. The s4 concept is discussed in Section 3 and Section 4 concludes the paper.

## 2 SELF-SUSTAINING SOFTWARE SYSTEMS (S4)

S4 supports the implementation of interpretable software systems capable of self-analysis and self-maintenance. Interpretability arises from collecting and managing the knowledge that defines modern software systems. This extended knowledge informs end-users, engineers, and the system itself. This section introduces the knowledge sources S4 intends to integrate and provides the first directions towards self-sustaining systems behaviour.

### 2.1 S4 Knowledge Sources

S4 integrates knowledge sources that expand the systems' knowledge base compared to current paradigms that only collect knowledge about systems' interactions. An extended knowledge base improves systems' interpretability and adaptability as more information is available to explain systems' behaviour and inform optimisation processes. Below, we introduce the knowledge sources and discuss the associated open research challenges.

*2.1.1 Systems Requirements and Design Artefacts.* The first knowledge source S4 considers comprises the end-user requirements and the systems' design artefacts. Engineers build systems according to the problems they solve. System designers carry out iterative elicitation processes that translate users' needs to requirements representing the stakeholders' expectations. Requirements drive the design of artefacts like functional and technical architectures and structural and behavioural models. These artefacts rule continuous processes of software development, testing, deployment, and maintenance [9]. Designers use modelling languages like Business Process Modelling and Notation (BPMN), the Unified Modelling Language (UML), or Systems Modeling Language (SysML) to represent end-user requirements and design artefacts. These models encapsulate knowledge about the problem systems address (i.e., systems goals) and the expertise of designers in architecting systems (i.e., designed solutions). S4 proposes to exploit this knowledge to enhance systems traceability, reusability, and interpretability [1], which improves the systems' understanding at deployment and enables adaptive decision processes that align with the systems' goals. However, this knowledge is hard to manage and update using current software design approaches. Organisations accumulate design models in architecture documents that are difficult to navigate [20] and do not have a direct link with the deployed systems [11].

S4 uses Knowledge Graphs (KGs) to model systems' requirements and software design decisions (i.e., the designers' expertise) in dynamic artefacts [20]. The idea is to create modular models linked to the deployed systems through data. KGs are semantic knowledge bases that describe the physical world [30]. The architecture of a KG encompasses a schema layer and a data layer [38]. The schema layer is the core of the knowledge graph that defines the conceptual meta-models based on ontologies [34]. This layer provides the representation capabilities that current software design modelling languages offer. Software design artefacts (e.g., the SysML requirements diagram) conform to ontologies or meta-models that define the modelling languages [19]. The data layer complements the schema layer and stores facts about the modelled entity [38]. We propose to exploit this data layer to close the gap between static software design artefacts and the dynamic nature of the evolving environment. A system at deployment produces data all the time, which can form the data layer of KG-based representations. Such a process enables the evolution of the design artefacts that reflect the system's changes. However, the exact methods behind this process are an open research challenge because it is hard to nurture KGs with the data produced by deployed software systems. Deployed systems usually follow architectures that hide data behind interfaces (e.g., microservices), which complicate access to it [3], and the translation of raw data into knowledge representations (i.e., data layer facts) is not straightforward. These challenges require further research to enable KGs as tools for software systems modelling.

*2.1.2 Deployed Systems.* S4 considers the deployed systems to be the second knowledge source. Current self-adaptive mechanisms monitor the state of internal components and input data to make decisions according to perceived changes [39]. However, data access tends to be localised, limiting the scope of the potential adaptation. Software systems could make better adaptive decision-making if they consider a holistic view of the entire data model and its states across multiple internal components. Modern software architectures encapsulate systems' functionalities into services that communicate through well-defined interfaces (i.e., APIs) [33]. These interfaces hide the systems' data and cause a situation known as 'The Data Dichotomy": while high-quality data management requires exposing systems' data, services hide it [32]. The Data Dichotomy impedes access to the systems' data and its posterior use in monitoring, transparency, traceability, and interpretability tasks. S4 leverages existing work on Data-Oriented Architecture (DOA) to overcome these limitations [3]. DOA complements the current paradigms for building and deploying systems. DOA considers data as the common denominator between disparate system components [15]. Services in DOA are distributed, autonomous, and communicate with each other at the data level (i.e., data coupling) using asynchronous messages [29, 37]. DOA enables systems to achieve data availability, reusability, and monitoring [15, 29, 37].

S4 pushes forward data-orientation ideas towards systems that make their data fully available and observable by design, facilitating systems monitoring and adaptation [3]. For example, a DOA-based system stores the states of the data that flows through it by design because of the data coupling between its components. A software agent or engineer can trace and analyse such records, identify when the system's data changes, and update the system's components. Existing work shows that building systems following DOA principles enables reasoning tasks at run-time (e.g., causality analysis [23]). S4 includes software systems designed following the DOA principles in the knowledge loop (Figure 1). DOA advocates for data-coupling between components to enable interpretability and better-informed self-adaptative decisions. Data coupling creates shared data models between systems' components, which store the whole systems' data and states, including internal and external interactions. Shared data models are interfaces with external entities like engineers or software agents that can monitor the system's status. This integration requires S4 to explore research areas for defining effective interfaces between the knowledge loop components and addressing data security and privacy concerns in DOAs. We propose to explore research efforts on Knowledge Graphs completion and construction [14, 25] to define effective interfaces between DOA-based systems and their representations and to include security mechanisms like homomorphic encryption [10] in DOA open and decentralised setups to address security and privacy challenges [3].

*2.1.3 Software Engineering Community.* Software engineering (SE) is a discipline that includes principles for documenting, developing, testing, deploying, and maintaining software systems [13]. The SE community has built a wealth of knowledge around all these processes. This knowledge is a valuable source of software architectural paradigms, design patterns, common pitfalls, and best practices, among other crucial information in understanding and maintaining software systems. This knowledge has been historically available in written documents (e.g., books, research papers, etc.) and online (e.g., collaborative knowledge bases such as Stack Overflow, GitHub or Reddit). Software engineers continuously use this knowledge to implement modern software systems. Rapid advances in Large Language Models (LLMs) open new opportunities to include this domain knowledge across the software systems life cycle. S4 proposes to develop domain-specific LLMs or leverage existing ones to encode the knowledge produced by the SE community and the domain knowledge of the problems engineers face and corresponding solutions. These models serve as an interface for engineers, software agents, and end-users to ask questions about the system (e.g., adaptation requests, behaviour explanations, etc.).

The use of LLMs in the SE context is growing. The community starts to understand the potential of this new technology but also its limitations [21]. Recent surveys show that most research efforts are focused on code generation, completion, testing, and fixing because such tasks benefit from the LLM's capability to generate code. Tasks like requirements analysis, systems improvement, and human-computer interaction require more attention in the future [8, 13]. S4 envisages a domain-specific LLM for each system, which adapts the general SE knowledge to the problems specific to this system. These domain-specific LLMs are an interface between systems, stakeholders (e.g., users and engineers) and autonomous agents. The stakeholders improve their understanding of a system as they can ask questions and get responses about its structure and behaviour in natural language. Autonomous agents can query the LLM to get enriched information and make better informed adaptive decisions. Building domain-specific LLMs presents significant challenges. Systems architectures, artefacts and their representations have unique properties that distinguish them from natural language, challenging the LLMs training [8]. LLMs generate different responses to the same question because of their non-deterministic nature. Such variability threatens robust, reliable, and stable systems' interpretations and adaptive processes [8]. There is also a risk of injecting bugs and replicating bad practices because of the LLMs hallucinations [8, 13, 21]. The environmental cost of training and deploying LLMs is a significant concern [21].

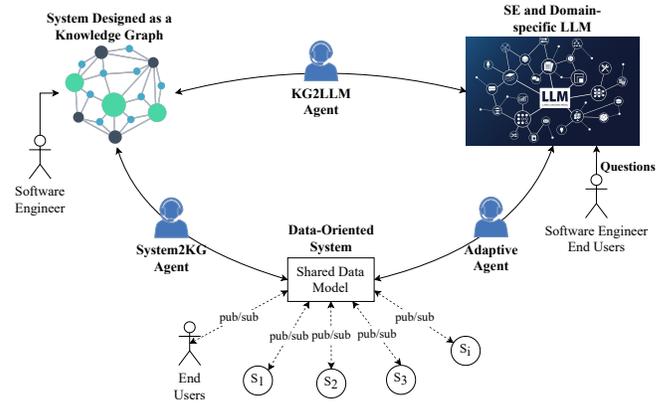

**Figure 1: S4 diagram depicting the knowledge sources that define software systems: Systems requirements and Design Artefacts, the Deployed Systems, and the SE Community. A knowledge loop integrates these sources through three autonomous agents to accomplish self-sustaining behaviour.**

## 2.2 Self-sustaining Behaviour

S4 aims to create systems with self-sustaining properties that improve their interpretability and adaptation capabilities. Different parts of S4 improve systems interpretability. A KG model is more flexible and easier to manage than traditional software design models. Stakeholders can use these models to understand the system structure [11, 27]. The shared data model of a DOA-based system stores the system's states through time. Such raw data is available but could be hard to read and understand. The domain-specific LLM offers a mechanism to address this limitation. The LLM is an interface for stakeholders to interact with the systems' design and data using natural language. S4 requires combining the knowledge sources described above to enable such interfaces. S4 takes inspiration from the multi-agent systems field [35] to integrate the knowledge sources (Figure 1) and to enable adaptation capabilities. We now describe the S4 autonomous agents.

The *System2KG Agent* connects the deployed system with its KG representation. This agent updates the KG system model with the system state by translating the data exposed in the shared data model to facts in the KG data layer. Traditional KGs provide static snapshots of knowledge structure, ignoring the evolving nature of knowledge. Evolving KGs expand their knowledge over time with continuously generated new facts [18]. This task is challenging and expensive [5]. The *System2KG Agent* builds on current research around evolving KGs by including the temporal dimension in the models' representations to perform tasks such as knowledge graph completion, entity discovery, and relation extraction [14]. The *KG2LLM Agent* bridges the updated KG system model with the domain-specific LLM. This agent injects structured information about the system's design and states into the LLM to improve response accuracy. The LLM provides feedback about the system's structure in the KG representation and supports its evolution. The *KG2LLM Agent* relies on current research exploring the enhancement of LLMs using KGs, the augmentation of KGs based on LLMs, and the synergy between the two. KGs enhance LLMs by including structured information in the LLM pre-training, inference, or interpretability stages. LLMs can augment KGs by constructing KG models from scratch, completing existing KGs, extracting knowledge from text, and supporting question-answering tasks. The synergic interaction between KGs and LLMs is the most promising, challenging and least explored research area. It includes using KGs and LLMs to build unified knowledge representations and applying them to reasoning in different applications [25].

The *Adaptive Agent* is responsible for the self-analysis and self-maintenance capabilities. This agent builds on the previous works around autonomous software development and systems to make adaptive decisions using the knowledge available from the combined sources. The agent monitors and analyses the KG system's representation and the shared data model to identify changes in the system's goals, design, or behaviour. It uses internal models and reasoning to make decisions that include the knowledge encoded in the domain-specific LLM. Finally, the agent executes the adaptive actions in the system and monitors their effect. The *Adaptive Agent* builds on existing automatic methods in the software development and deployment processes rather than considering the LLM an oracle that makes the entire decision-making. For example, software testing has a significant degree of automation, which S4 extends by generating unit tests that improve testing coverage [21]. Genetic improvement and self-adaptive systems approaches improve systems' performance regarding simple non-functional requirements (e.g., accuracy, response time, etc.) [2, 17]. S4 expands such capabilities by including more knowledge in the decision-making process.

## 3 DISCUSSION

S4 proposes a new research agenda to address the limitations of current paradigms for building interpretable adaptive systems capable of self-analysis and self-maintenance. The research agenda includes developing new approaches for formalising and managing the knowledge sources that define modern software systems: systems requirements and design artefacts modelled as KGs, the deployed systems designed following DOA principles, and the SE community knowledge encoded in LLMs. A knowledge loop combines these sources to improve the systems' interpretability and adaptation. Such adaptive processes include the internal systems structure (i.e., software source code), the system's behaviour, and their environments' configuration. S4 improves interpretability by offering natural language interfaces between stakeholders and software systems based on the available knowledge.

The research agenda of S4 is ambitious and aims to transform SE practices leveraging advances in semantic understanding of software, data-oriented architectures, and foundational models. Such transformations are challenging to carry out. For example, designing software systems with evolving KG models is an open research area. The main challenges emerge from the technical difficulty of translating raw data into knowledge representations [5]. DOA-based systems also have challenges regarding the privacy and security of data in the proposed open and decentralised setups [3]. LLMs are non-deterministic and prone to hallucination, which complicates their adoption in tasks that require robust and reliable responses (e.g., critical systems adaptation) [8, 13, 21]. The S4 research agenda builds on existing research in related fields towards enabling the required shifts. We believe interdisciplinary research projects will be crucial to exploiting the expertise developed by different research communities.

The success of the S4 research agenda depends on the software engineering and systems community embracing the self-sustaining concept. We aim to initiate a discussion on current autonomous computing paradigms and potential alternatives through this position paper. Our future work includes developing prototypes to evaluate the presented ideas and translating these prototypes to real-world systems by working with industry partners and practitioners in different domains. Such development and validation processes will enable community awareness around the opportunities, strengths, and limitations of the S4 research agenda.

## 4 CONCLUSIONS

This position paper introduces the idea of Self-Sustaining Software Systems (S4) as an alternative paradigm to design and deploy interpretable autonomous systems. S4 integrates the knowledge sources that define modern software systems: the systems requirements and design artefacts, the deployed system, and the SE community. A set of agents integrates these knowledge sources in a knowledge loop to improve the system's self-awareness, allow for automated maintenance, and allow stakeholders to interact with the system using natural language. We highlight the open research challenges associated with the S4 idea and discuss the threats towards its vision realisation. Our future work will focus on developing the S4 idea with incremental prototypes and evaluating them in different real-world domains.

## REFERENCES
[1] Audrey Berquand and Annalisa Riccardi. 2020. From engineering models to knowledge graph: delivering new insights into models. In *"9th International Systems & Concurrent Engineering for Space Applications Conference (SECESA 2020)"*.
[2] Christian Cabrera and Siobhán Clarke. 2022. A Self-Adaptive Service Discovery Model for Smart Cities. *IEEE Transactions on Services Computing* 15, 1 (2022), 386–399. https://doi.org/10.1109/TSC.2019.2944356
[3] Christian Cabrera, Andrei Paleyes, Pierre Thodoroff, and Neil D Lawrence. 2023. Real-world Machine Learning Systems: A survey from a Data-Oriented Architecture Perspective. *arXiv preprint arXiv:2302.04810* (2023).
[4] Matteo Camilli, Raffaela Mirandola, and Patrizia Scandurra. 2023. Enforcing Resilience in Cyber-Physical Systems via Equilibrium Verification at Runtime. *ACM Trans. Auton. Adapt. Syst.* 18, 3 (sep 2023), 32 pages. https://doi.org/10.1145/3584364

[5] Sutanay Choudhury, Khushbu Agarwal, Sumit Purohit, Baichuan Zhang, Meg Pirrung, Will Smith, and Mathew Thomas. 2017. NOUS: Construction and Querying of Dynamic Knowledge Graphs. In *2017 IEEE 33rd International Conference on Data Engineering (ICDE)*. 1563–1565. https://doi.org/10.1109/ICDE.2017.228

[6] Jennifer Cobbe, Michael Veale, and Jatinder Singh. 2023. Understanding Accountability in Algorithmic Supply Chains. In *Proceedings of the 2023 ACM Conference on Fairness, Accountability, and Transparency* (Chicago, IL, USA) *(FAccT '23)*. Association for Computing Machinery, New York, NY, USA, 1186–1197. https://doi.org/10.1145/3593013.3594073

[7] Agoston E Eiben and James E Smith. 2015. *Introduction to evolutionary computing*. Springer.

[8] Angela Fan, Beliz Gokkaya, Mark Harman, Mitya Lyubarskiy, Shubho Sengupta, Shin Yoo, and Jie M. Zhang. 2023. Large Language Models for Software Engineering: Survey and Open Problems. *arXiv preprint arXiv:2310.03533* (2023).

[9] Brian Fitzgerald and Klaas-Jan Stol. 2017. Continuous software engineering: A roadmap and agenda. *Journal of Systems and Software* 123 (2017), 176–189. https://doi.org/10.1016/j.jss.2015.06.063

[10] Caroline Fontaine and Fabien Galand. 2007. A survey of homomorphic encryption for nonspecialists. *EURASIP Journal on Information Security* 2007 (2007), 1–10.

[11] Chao Fu, Jihong Liu, Longxi Zhang, Yinxuan Mao, and Jie Jin. 2021. Knowledge graph based System model configuration design. *Journal of Physics: Conference Series* 2029, 1 (sep 2021), 012108. https://doi.org/10.1088/1742-6596/2029/1/012108

[12] Yu Gan, Yanqi Zhang, Dailun Cheng, Ankitha Shetty, Priyal Rathi, Nayan Katarki, Ariana Bruno, Justin Hu, Brian Ritchken, Brendon Jackson, Kelvin Hu, Meghna Pancholi, Yuan He, Brett Clancy, Chris Colen, Fukang Wen, Catherine Leung, Siyuan Wang, Leon Zaruvinsky, Mateo Espinosa, Rick Lin, Zhongling Liu, Jake Padilla, and Christina Delimitrou. 2019. An Open-Source Benchmark Suite for Microservices and Their Hardware-Software Implications for Cloud & Edge Systems. In *Proceedings of the Twenty-Fourth International Conference on Architectural Support for Programming Languages and Operating Systems* (Providence, RI, USA) *(ASPLOS '19)*. Association for Computing Machinery, New York, NY, USA, 3–18. https://doi.org/10.1145/3297858.3304013

[13] Xinyi Hou, Yanjie Zhao, Yue Liu, Zhou Yang, Kailong Wang, Li Li, Xiapu Luo, David Lo, John Grundy, and Haoyu Wang. 2023. Large Language Models for Software Engineering: A Systematic Literature Review. *arXiv preprint arXiv:2308.10620* (2023).

[14] Shaoxiong Ji, Shirui Pan, Erik Cambria, Pekka Marttinen, and Philip S. Yu. 2022. A Survey on Knowledge Graphs: Representation, Acquisition, and Applications. *IEEE Transactions on Neural Networks and Learning Systems* 33, 2 (2022), 494–514. https://doi.org/10.1109/TNNLS.2021.3070843

[15] Rajive Joshi. 2007. Data-oriented architecture: A loosely-coupled real-time SOA. *whitepaper, Aug* (2007).

[16] Philippe Lalanda, Julie A McCann, and Ada Diaconescu. 2013. *Autonomic computing: principles, design and implementation*. Springer Science & Business Media.

[17] William B. Langdon and Mark Harman. 2015. Optimizing Existing Software With Genetic Programming. *IEEE Transactions on Evolutionary Computation* 19, 1 (2015), 118–135. https://doi.org/10.1109/TEVC.2013.2281544

[18] Jiaqi Liu, Qin Zhang, Luoyi Fu, Xinbing Wang, and Songwu Lu. 2019. Evolving Knowledge Graphs. In *IEEE INFOCOM 2019 - IEEE Conference on Computer Communications*. 2260–2268. https://doi.org/10.1109/INFOCOM.2019.8737547

[19] Chase McCoy. [n. d.]. *Ontology Action Team - MBSE Wiki*. https://www.omgwiki.org/MBSE/doku.php?id=mbse:ontology

[20] Chase McCoy. 2021. *Design Systems as Knowledge Graph*. https://chasem.co/2021/08/systems-as-knowledge-graphs

[21] Ipek Ozkaya. 2023. Application of Large Language Models to Software Engineering Tasks: Opportunities, Risks, and Implications. *IEEE Software* 40, 3 (2023), 4–8. https://doi.org/10.1109/MS.2023.3248401

[22] Claus Pahl. 2023. Research challenges for machine learning-constructed software. *Service Oriented Computing and Applications* 17, 1 (2023), 1–4.

[23] Andrei Paleyes, Siyuan Guo, Bernhard Scholkopf, and Neil D. Lawrence. 2023. Dataflow graphs as complete causal graphs. In *2023 IEEE/ACM 2nd International Conference on AI Engineering – Software Engineering for AI (CAIN)*. 7–12. https://doi.org/10.1109/CAIN58948.2023.00010

[24] Andrei Paleyes, Raoul-Gabriel Urma, and Neil D. Lawrence. 2022. Challenges in Deploying Machine Learning: A Survey of Case Studies. *ACM Comput. Surv.* (apr 2022). https://doi.org/10.1145/3533378

[25] Shirui Pan, Linhao Luo, Yufei Wang, Chen Chen, Jiapu Wang, and Xindong Wu. 2023. Unifying Large Language Models and Knowledge Graphs: A Roadmap. *arXiv preprint arXiv:2306.08302* (2023).

[26] Justyna Petke, Saemundur O. Haraldsson, Mark Harman, William B. Langdon, David R. White, and John R. Woodward. 2018. Genetic Improvement of Software: A Comprehensive Survey. *IEEE Transactions on Evolutionary Computation* 22, 3 (2018), 415–432. https://doi.org/10.1109/TEVC.2017.2693219

[27] Satrio Adi Rukmono and Michel R.V. Chaudron. 2023. Enabling Analysis and Reasoning on Software Systems through Knowledge Graph Representation. In *2023 IEEE/ACM 20th International Conference on Mining Software Repositories (MSR)*. 120–124. https://doi.org/10.1109/MSR59073.2023.00029

[28] Ina Schieferdecker. 2020. *Responsible Software Engineering*. Springer International Publishing, Cham, 137–146. https://doi.org/10.1007/978-3-030-29509-7_11

[29] Robert Schuler, Carl Kesselman, and Karl Czajkowski. 2015. Data Centric Discovery with a Data-Oriented Architecture. In *Proceedings of the 1st Workshop on The Science of Cyberinfrastructure: Research, Experience, Applications and Models* (Portland, Oregon, USA) *(SCREAM '15)*. Association for Computing Machinery, New York, NY, USA, 37–44. https://doi.org/10.1145/2753524.2753532

[30] Amit Sheth, Swati Padhee, and Amelie Gyrard. 2019. Knowledge Graphs and Knowledge Networks: The Story in Brief. *IEEE Internet Computing* 23, 4 (2019), 67–75. https://doi.org/10.1109/MIC.2019.2928449

[31] Jatinder Singh, Jennifer Cobbe, and Chris Norval. 2019. Decision Provenance: Harnessing Data Flow for Accountable Systems. *IEEE Access* 7 (2019), 6562–6574. https://doi.org/10.1109/ACCESS.2018.2887201

[32] Ben Stopford. 2016. *The Data Dichotomy: Rethinking the Way We Treat Data and Services*. Available at https://www.confluent.io/blog/data-dichotomy-rethinking-the-way-we-treat-data-and-services/.

[33] Davide Taibi, Valentina Lenarduzzi, and Claus Pahl. 2018. Architectural patterns for microservices: A systematic mapping study. In *CLOSER 2018 - Proceedings of the 8th International Conference on Cloud Computing and Services Science*. SCITEPRESS, 221–232. https://doi.org/10.5220/0006798302210232 International Conference on Cloud Computing and Services Science ; Conference date: 19-03-2018 Through 21-03-2018.

[34] Steve Tueno, Régine Laleau, Amel Mammar, and Marc Frappier. 2017. Towards Using Ontologies for Domain Modeling within the SysML/KAOS Approach. In *2017 IEEE 25th International Requirements Engineering Conference Workshops (REW)*. 1–5. https://doi.org/10.1109/REW.2017.22

[35] Adelinde M Uhrmacher and Danny Weyns. 2009. *Multi-Agent systems: Simulation and applications*. CRC press.

[36] Ashish Vaswani, Noam Shazeer, Niki Parmar, Jakob Uszkoreit, Llion Jones, Aidan N Gomez, Ł ukasz Kaiser, and Illia Polosukhin. 2017. Attention is All you Need. In *Advances in Neural Information Processing Systems*, I. Guyon, U. Von Luxburg, S. Bengio, H. Wallach, R. Fergus, S. Vishwanathan, and R. Garnett (Eds.), Vol. 30. Curran Associates, Inc. https://proceedings.neurips.cc/paper_files/paper/2017/file/3f5ee243547dee91fbd053c1c4a845aa-Paper.pdf

[37] Christian Vorhemus and Erich Schikuta. 2017. A data-oriented architecture for loosely coupled real-time information systems. In *Proceedings of the 19th International Conference on Information Integration and Web-based Applications & Services*. 472–481.

[38] Lu Wang, Chenhan Sun, Chongyang Zhang, Weikun Nie, and Kaiyuan Huang. 2023. Application of knowledge graph in software engineering field: A systematic literature review. *Information and Software Technology* 164 (2023), 107327. https://doi.org/10.1016/j.infsof.2023.107327

[39] Danny Weyns. 2019. Software engineering of self-adaptive systems. *Handbook of software engineering* (2019), 399–443.

[40] Terence Wong, Markus Wagner, and Christoph Treude. 2022. Self-adaptive systems: A systematic literature review across categories and domains. *Information and Software Technology* 148 (2022), 106934. https://doi.org/10.1016/j.infsof.2022.106934